\documentclass[preprintnumbers,amsmath,amssymb,superscriptaddress,twocolumn]{revtex4}
\usepackage{dcolumn}
\usepackage{amsmath}
\usepackage{color}
\usepackage{graphicx}
\usepackage{epstopdf}
\usepackage{ulem}
\usepackage{float}

\begin{document}
\title{Photon number discrimination without a photon counter and its application to reconstructing non-Gaussian states}

\author{H.~M.~Chrzanowski}
\affiliation{Centre for Quantum Computation and Communication Technology} 
\affiliation{Quantum Optics group,  Department of Quantum Science, Research School of Physics and Engineering, Australian National University, Canberra ACT 0200, Australia.}
\author{J.~Bernu}
\affiliation{Centre for Quantum Computation and Communication Technology} 
\affiliation{Quantum Optics group,  Department of Quantum Science, Research School of Physics and Engineering, Australian National University, Canberra ACT 0200, Australia.}
\author{B.~M.~Sparkes}
\affiliation{Centre for Quantum Computation and Communication Technology} 
\affiliation{Quantum Optics group,  Department of Quantum Science, Research School of Physics and Engineering, Australian National University, Canberra ACT 0200, Australia.}
\author{B.~Hage}
\affiliation{Centre for Quantum Computation and Communication Technology} 
\affiliation{Quantum Optics group,  Department of Quantum Science, Research School of Physics and Engineering, Australian National University, Canberra ACT 0200, Australia.}
\author{A.~P.~Lund}
\affiliation{Centre for Quantum Computation and Communication Technology} 
\affiliation{Centre for Quantum Dynamics, Griffith University, Nathan QLD 4111, Australia.}
\author{T.~C.~Ralph}
\affiliation{Centre for Quantum Computation and Communication Technology} 
\affiliation{Department of Physics, University of Queensland, St. Lucia QLD 4072, Australia.}
\author{P.~K.~Lam}
\affiliation{Centre for Quantum Computation and Communication Technology} 
\affiliation{Quantum Optics group,  Department of Quantum Science, Research School of Physics and Engineering, Australian National University, Canberra ACT 0200, Australia.}
\author{T.~Symul}
\affiliation{Centre for Quantum Computation and Communication Technology} 
\affiliation{Quantum Optics group,  Department of Quantum Science, Research School of Physics and Engineering, Australian National University, Canberra ACT 0200, Australia.}

\newcommand{\eg}{\textit{e.g.} }
\newcommand{\Xb}{\hat X^\theta_b}
\newcommand{\na}{\hat n_a}

\begin{abstract}
The non-linearity of a conditional photon-counting measurement can be used to `de-Gaussify' a Gaussian state of light. Here we present and experimentally demonstrate a technique for photon number resolution using only homodyne detection. We then apply this technique to inform a conditional measurement; unambiguously reconstructing the statistics of the non-Gaussian one and two photon subtracted squeezed vacuum states. Although our photon number measurement relies on ensemble averages and cannot be used to prepare non-Gaussian states of light, its high efficiency, photon number resolving capabilities, and compatibility with the telecommunications band make it suitable for quantum information tasks relying on the outcomes of mean values. 
\end{abstract}

\maketitle

Historically, quantum optics has been divided into two complementary camps, each exploiting only one aspect of the wave-particle duality of light. The ``Discrete Variable" (DV) approach relies on quantized measurements of optical systems with low photon numbers, where the qubit space is often spanned by two orthogonal polarizations. In contrast, the ``Continuous Variable" (CV) approach focuses on field measurements of comparatively bright beams, with states defined in an infinite-dimensional Hilbert space. With quantum optics providing an ideal field for the first demonstrations of quantum information protocols, non-Gaussian states and transformations are a necessity, forming a crucial resource for quantum communications~\cite{COC98,OUR09}, metrology~\cite{RAL02,gilchrist} and quantum computing~\cite{ralph2, jeong, LUN08}. CV techniques are popular candidate quantum information protocols, owing to their high detection efficiency and compatibility with existing telecommunication infrastructures. However, the usual continuous variable toolbox, comprising Gaussian states, linear optics and homodyne detection, is insufficient to break out of the Gaussian regime.

In the absence of extreme nonlinearities, integrating DV photon-counting into continuous variable setups has been shown to provide probabilistic `de-Gaussification'~\cite{ourjoumtsev3,ourjoumtsev1,neergaardnielsen,wakui,ourjoumtsev2,parigi,namekata, gerrit}. This approach circumvents the `Gaussian' limitations of CV whilst enjoying the degrees of freedom both techniques provide. However, these `hybrid' experiments face challenges arising from simultaneously exploiting both the wave and particle properties of light. In this paper we demonstrate a continuous variable analog to the `photon counter' that enables discrimination of quanta of light. We show the versatility and efficacy of this technique by experimentally reconstructing the non-Gaussian one and two photon subtracted squeezed vacuum (1 and 2-PSSV) states.

CV techniques combined with linear optics are known to be insufficient to prepare non-Gaussian states with negativity in their Wigner functions \cite{giedke}.  Nevertheless, the idea of measuring the corpuscular nature of light with only CV techniques has been theoretically~\cite{Leonhardt2,RAL00,ralph1} and experimentally~\cite{vasilyev,WEB06,GRO07}  investigated. Here we extend these ideas and show how DV heralding can be replaced by pure CV conditioning  for the reconstruction of non-Gaussian states. This protocol avoids experimental issues arising from `hybridising' a setup~\cite{namekata} whilst harnessing the existing mode selectivity, high quantum efficiency and low dark noise of homodyne detection. Although the $k$-PSSV states are not heralded, remarkably we can still extract their quantum statistics. Using this method, we have successfully reconstructed the 1 and 2 PSSV states.

\begin{figure}[!ht]
\centering
\includegraphics[width=8.5cm]{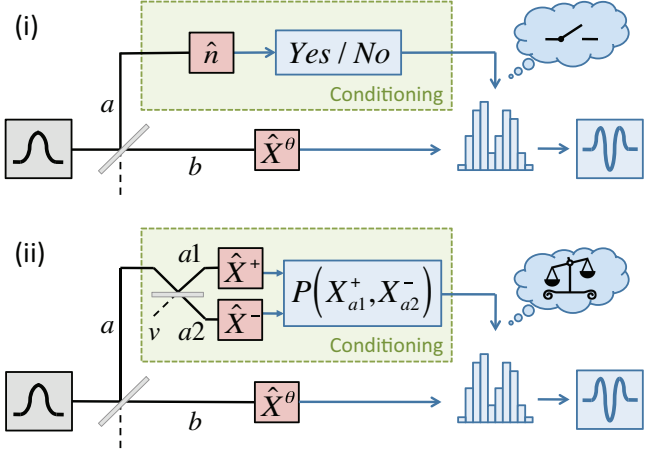}
\caption{ (Color online) \textbf{Schematic of $k$-photon subtracted squeezed vacuum states reconstruction.} Both schemes require squeezed Gaussian input states.  A beam-splitter is used to tap-off a small portion of the input beam, $a$, for conditioning measurement.  The remaining light, $b$, is analysed with tomographic measurements.  \textbf{(i)~Hybrid Setup}: outcomes from photon number resolving detectors are used to gate the results of the tomographic detection.  Keeping only the conditionally heralded statistics, non-Gaussian quantum states are reconstructed;  \textbf{(ii)~Pure Continuous Variable Setup}: simultaneous orthogonal quadrature measurements are performed to weight the statistics of the tomographic measurements, producing non-Gaussian statistics.}
\label{figure1}
\end{figure}

\begin{figure}
\centering
\includegraphics[width=9cm]{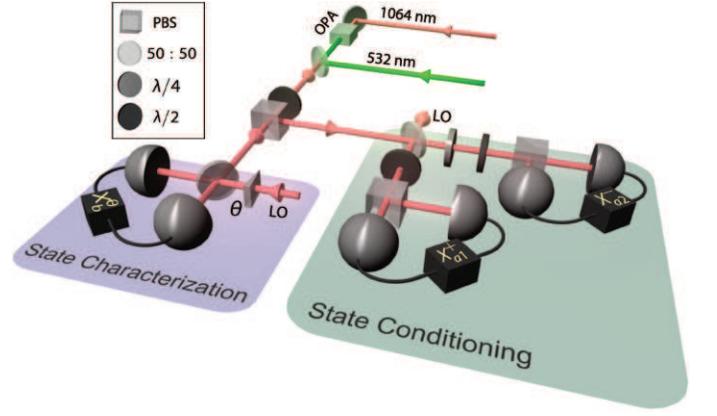}
\caption{(Color online) \textbf{Experimental set-up for the pure CV protocol.} A squeezed vacuum state ($V_s=-3.78$ dB of squeezing for $V_a=+4.33$ dB of antisqueezing)  at the sideband frequencies between 3.6-4.2 MHz of a bright carrier is produced by a degenerate optical parametric amplifier (OPA).
A variable beam splitter, a $\lambda / 2$ wave plate combined with a polarising beam splitter (PBS), reflects 20\% of the incoming beam for conditioning and transmits the remaining for tomographic reconstruction. The conditioning beam is split between two homodyne detections (channels $a_1$ and $a_2$) measuring two arbitrary orthogonal quadratures $X^\phi_{a1}$ and $X^{\phi+\frac{\pi}{2}}_{a2}$. The orthogonality between the homodyning angles for modes $a1$ and $a2$ is set using a combination of polarisation optics, whilst the global phase $\phi$ does not require active control (see eqn.\ref{ntoXY}). The tomographic reconstruction is preformed by sampling $X^{\theta}_b$ at 12 fixed angles $\theta$ between $0$ and $180^{\circ}$ in intervals of $15^\circ$. The experiment is controlled and automated using an FPGA based system ~\cite{sparkes}. }
\label{figure2}
\end{figure}

\begin{figure}
\centering
\includegraphics[width=8cm]{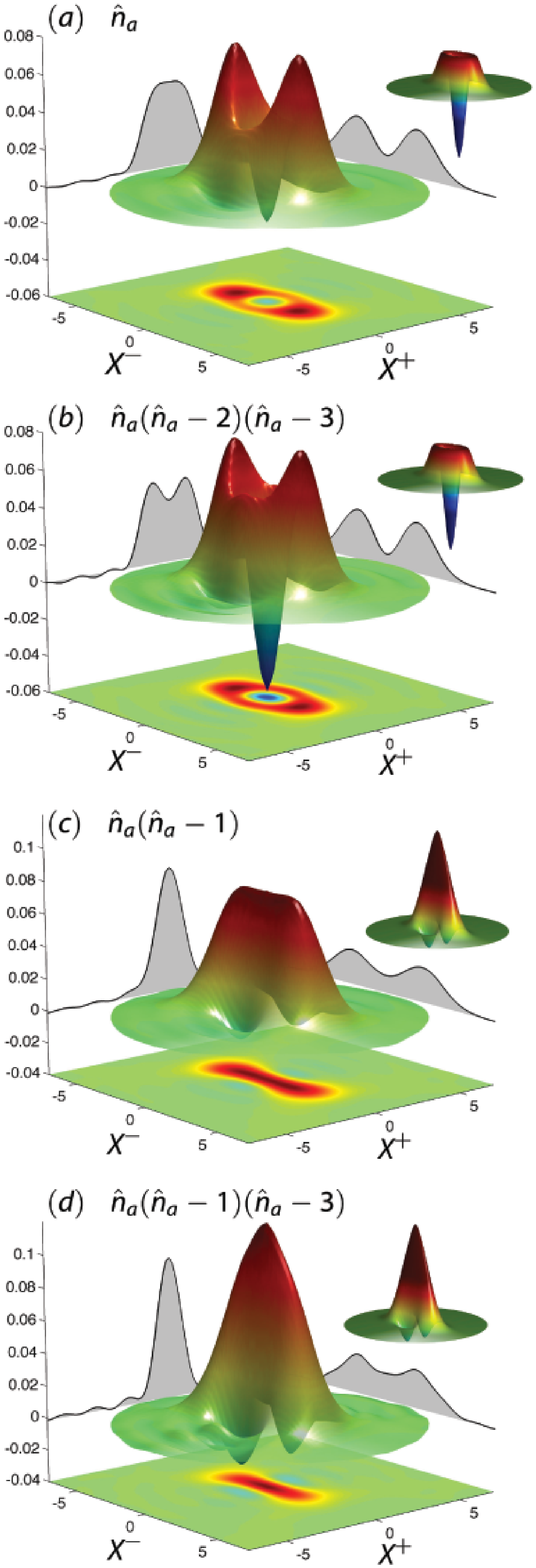}
\caption{(Color online) Wigner Functions reconstructed via the inverse Radon transform. The specific weighting polynomial is given at the top of each diagram. \textbf{1-PSSV}: \textbf{(a)}~without and \textbf{(b)}~with higher order correction respectively, showing a improvement in the negativity from $-0.018$ to $-0.055$ compared to $-0.16$ in the ideal case. \textbf{2-PSSV}: \textbf{(c)}~without and \textbf{(d)}~with higher order correction respectively. It shows an improvement in the size of the central positive fringes (and first negative fringe) from $+0.077$ ($-0.011$) to $+0.122$ ($-0.023$) compared to $+0.16$ ($-0.03$) in the ideal case. The insets display the corresponding calculated 1 and 2-PSSV states assuming pure initial squeezed states and ideal photon subtraction. The shadows represent the reconstructed quadrature distributions, obtained from the marginals of the reconstructed Wigner functions.}
\label{figure3}
\end{figure} 

A beam splitter is used to divert a small portion of an input squeezed vacuum (mode $a$) for conditioning whilst the remainder (mode $b$) is sent to a tomographic homodyne detection that measures $\hat{X}^{\theta}_b=e^{-\imath \theta}\hat{a}_b+e^{\imath \theta}\hat{a}_b^\dagger$, where $\hat{a}_b$ and $\hat{a}_b^\dagger$ are the annihilation and creation operators in mode $b$ and $\theta$ is the quadrature angle. The probability distributions of the $k$-PSSV are given by
\begin{equation}
\pi^\theta_k(x)\equiv \pi(X^\theta_b=x|n_a=k).
\label{prob}
\end{equation}
In hybrid experiments (see Fig.\ref{figure1}(i)) these are simply estimated by reconstructing histograms of $X^{\theta}_b$ only when $k$ photons are detected in mode $a$. The CV-only approach presented here (see Fig.\ref{figure1}(ii)) replaces the DV conditioning with a dual-homodyne measurement and exploits the relationship between photon number and field operators:
\begin{equation}
\hat{n}_a= \hat{a}^\dagger_a \hat{a}_a=\frac{1}{4} \left[(\hat{X}^{+}_a)^2+(\hat{X}^{-}_a)^{2}-2\right],
\label{ntoXY}
\end{equation}
where $\hat{X}^+_a\!=\!\hat{X}^{\phi=0}_a$ and $\hat{X}^{-}_a\!=\!\hat{X}^{\phi=\pi/2 }_a$, or any other pair of orthogonal quadratures $\hat{X}^{\phi}_a$ and $\hat{X}^{\phi+\pi/2}_a$.
Perfect simultaneous measurement of conjugate observables $X^+_a$ and $X^-_a$ being forbidden, the two observables $ X^+_{a1}$ and $X^-_{a2}$ sampled by the dual homodyne detection incur a noise penalty in the form of vacuum fluctuations entering the empty input port $v$ of the beam-splitter. (see Fig.\ref{figure1}(ii)):
\begin{equation}
\begin{array}{ccc}
\hat X^+_{a1} & = & (\hat X^+_a+\hat X^+_v)/\sqrt 2 \\
\hat X^-_{a2} & = & (\hat X^-_a-\hat X^-_v)/\sqrt 2.
\end{array}
\label{BS}
\end{equation}
As a consequence, attempts to measure $n_a$ by using dual homodyne detection produce continuous real values rather than integer results. This prevents any direct heralding of the $k$-PSSV with Gaussian measurements only, in accordance with \cite{giedke}, as exploiting the measurement outcome for single shot heralding would lead to meaningless results. However, since a tomographic reconstruction only deals with ensemble measurements, and because the statistical properties of the vacuum fluctuations are perfectly known, this additional noise can be effectively cancelled.
We first rewrite $\pi^\theta_k(x)$ as
\begin{equation}
\pi^\theta_k(x)=\frac{ \pi(X^\theta_b=x,n_a=k)}{\pi(n_a=k)}=\frac{\langle\delta(\hat X^\theta_b-x)\delta_{\hat n_ak}\rangle}{\pi(n_a=k)}.
\label{prob2}
\end{equation}
We then note that $\delta_{\hat n_ak}$ can be approximated by a polynomial $P_k(\hat n_a)$ equal to zero at any significant photon number other than $k$, for example $P_k(\hat n_a)=\hat n_a(\hat n_a-1)...(\hat n_a-k+1)$ in the limit of low squeezing. Dropping the normalisation factor, we find 
\begin{equation}
\pi^\theta_k(x)\simeq\langle\delta(\hat X^\theta_b-x)P_k(\hat n_a)\rangle.
\label{prob3}
\end{equation}
This means $\pi^\theta_k(x)$ can be experimentally obtained by constructing histograms of $X^\theta_b$ where \textit{all} values are taken but \textit{weighted} by the corresponding value of $P(n_a)$, effectively discarding results corresponding to unwanted $n_a\ne k$. In view to reconstructing the statistics of the 1-PSSV state, the simplest applied weighting, where $ P(n_a) = n_a$ can be understood as eliminating contributions of $X_b^{\theta}$ corresponding to $n_{a}=0$ (\textit{i.e.} no photon subtraction), and retaining the desired contributions where $n_{a}=1$ (\textit{i.e.} successful photon subtraction). 
This trivial remark becomes powerful when injecting (\ref{ntoXY}) in (\ref{prob3}), thus defining $P(\hat X^+_a,\hat X^-_a)$: we can now use the real valued outcomes of our dual homodyne detection to weigh the results of our tomographic measurements. To do so we need to transform $P(\hat X^+_a,\hat X^-_a)$ into another polynomial $Q(\hat X^+_{a1,}\hat X^-_{a2})$ of the measured observables that performs, on average, the same weighting. 
$Q(\hat X^+_{a1},\hat X^-_{a2})$, in general, is not trivial to obtain due to non~commuting algebra. The demonstration of the existence of $Q$ for any $P(\na)$, and algorithms to compute it, will be published elsewhere. Here, as an illustration of the concepts used in the general case, we only prove the result $Q(\hat X^+_{a1},\hat X^-_{a2})=[(\hat{X}^{+}_{a1})^2+(\hat{X}^{-}_{a2})^{2}-2]/2$ for the simpler case $P(\hat n_a)=\hat n_a$.
Using (\ref{ntoXY}) and (\ref{BS}) we find
\begin{equation}
\begin{array}{l}
\left\langle\delta(\Xb-x) Q(\hat X^+_{a1},\hat X^-_{a2})\right\rangle=\\
\hspace{1.5cm}\left\langle\delta(\Xb-x) \hat n_{a}\right\rangle +\left\langle\delta(\Xb-x) \hat \Delta\right\rangle
\end{array}
\label{delta}
\end{equation}
where
\begin{eqnarray}
\hat \Delta \hphantom{^\pm}& = & (\hat \Delta^++\hat \Delta^-)/2\\
\hat \Delta^\pm & = & (\hat X^\pm_v)^2-1\pm\hat X^\pm_a\hat X^\pm_v\pm\hat X^\pm_v\hat X^\pm_a.
\end{eqnarray}
We now note that the vacuum fluctuations $\hat X^\pm_v$ are uncorrelated to those of $\hat X^\pm_a$ and $\Xb$ and hence can be averaged out separately in (\ref{delta}), for example $\langle\delta(\Xb-x) \hat X^+_a\hat X^+_v\rangle=\langle\delta(\Xb-x)\hat X^+_a\rangle\langle\hat X^+_v\rangle$. Finally, injecting the values $\langle\hat X^\pm_v\rangle=0$ and $\langle (\hat X^\pm_v)^2\rangle=1$ given by quantum theory, we easily find that $\langle\delta(\Xb-x) \hat \Delta\rangle=0$. Hence we can experimentally obtain $\pi^\theta_k(x)$ through weighted histograms:
\begin{equation}
\pi^\theta_k(x) = \left\langle\delta(\Xb-x) Q(\hat X^+_{a1},\hat X^-_{a2})\right\rangle.
\end{equation}

As  hinted at earlier, this idea can be extended to more sophisticated polynomials $P(\hat{n}_a)$ that approach photon number selectivity, permitting us to reconstruct purer and/or larger $k$-PSSV states. Larger $k$-PSSV states can be reconstructed using polynomials that remove the contributions associated with subtracted photon numbers smaller than $k$, and keep outcomes corresponding to $n_a=k$ as the predominant contribution. For example, the 2-PSSV state can be reconstructed using $P(\hat{n}_a)=\hat{n}_a(\hat{n}_a-1)$, removing contributions resulting from having 0 or 1 photon in mode $a$.

We can also apply the same technique to remove contributions corresponding to unwanted higher order photon number subtractions contaminating the weighted ensemble. If we consider the simplest $P(\hat{n}_a)=\hat{n}_{a}$ conditioning polynomial to reconstruct the $1$-PSSV, whilst contributions corresponding to $n_a = 0$ are cancelled, measurements associated with two photon subtraction are kept, and their statistical contribution is weighted at $2$. As a result the reconstructed state is a statistical mixture of the $1$-PSSV with some contribution from the $2$-PSSV, producing a partial wash-out of the negativity of the reconstructed Wigner function. However, using $P(\hat{n}_a)=\hat{n}_a(\hat{n}_a-2)$ for conditioning allows us to remove contributions resulting from having $0$ or $2$ photons in mode $a$. These two conditioning techniques can, in theory, be extended to an arbitrary order allowing a CV analog to the DV photon number resolving detector.

Our experimental setup is detailed in Fig.\ref{figure2}. All of the presented Wigner functions (Fig.\ref{figure3}) are reconstructed directly from the probability distributions obtained by applying various conditioning polynomials to a unique dataset. This dataset is composed of approximately 1.2 billion samples for each detection mode and for each of the 12 tomographic angles. Whilst some reconstruction methods require assumptions on the nature of the state, the inverse Radon transform~\cite{Leonhardt} used here is direct and assumption free. Moreover we do not correct for any experimental inefficiencies.
 
Fig.\ref{figure3}(a) shows the Wigner function obtained using the conditioning polynomial $P(\na)=\hat{n}_a$, and displays a clear negativity. Our protocol, however, relies on correlations between modes $a1$, $a2$ and $b$, which are affected by any process that adds extra uncorrelated classical or quantum fluctuations. The primary limitation is the finite purity of our squeezed vacuum ressource, that can be modelled as an effective loss of $\sim\!\!\!12\%$ applied on a pure squeezed state. We note that this non-ideal resource is also a limiting factor in hybrid experiments~\cite{namekata}. Another possible limitation comes from the finite homodyne efficiency and dark noise. The dark noise, analogous to the dark counts of DV detectors, is at least 22 dB smaller than the vacuum fluctuations. Our homodyne efficiency is limited by the quantum efficiency of the detectors, estimated at between 93-96\%, whilst mode matching efficiency is typically greater than 99\%. In contrast, hybrid experiments suffer a substantial loss contribution from the difficulty in isolating the correct spatio-temporal mode in the conditioning and characterisation stages. Finally, the need for a finite conditioning tap-off and squeezing inevitably introduces spurious higher order photon subtraction contributions in the reconstructed state. By using the conditioning polynomial $P(\hat{n}_a)=\hat{n}_a(\hat{n}_a-2)(\hat{n}_a-3)$, the contributions from 2 and 3 photon subtractions are removed; the further higher orders being negligible. As expected, Fig.\ref{figure3}(b) shows a considerable improvement in the negativity of the reconstructed $1$-PSSV state.

Fig.\ref{figure3}(c)  shows the $2$-PSSV obtained using $P(\hat{n}_a)=\hat{n}_a(\hat{n}_a-1)$. The reconstructed state exhibits the expected central positive fringe and two negative side fringes, as well as a bigger separation of the two coherent components~\cite{dakna}. As shown in Fig.\ref{figure3}(d), correcting for higher order contaminations by using $P(\hat{n}_a)=\hat{n}_a(\hat{n}_a-1)(\hat{n}_a-3)$ enhances the purity of the $2$-PSSV state, evidenced by an improvement in the size of the fringes.

In this paper we have demonstrated a photon discrimination technique based on a dual homodyne detection, and used it to reconstruct conditional non-Gaussian states using only Gaussian resources and measurements and linear optics. Whilst this technique does not allow us to directly prepare non-Gaussian states, it enables their characterisation, and can be extended to characterise other quantum information protocols relying on hybrid, or `de-Gaussification`, techniques. By avoiding direct photon-counting, we circumvented the difficulties arising from simultaneously exploiting the wave and particle nature of light, permitting us to unambiguously reconstruct the $1$ and $2$-PSSV states. This idea of probing the `quantised' nature of the quantum system via probing of its continuous variables could also prove interesting for fields such as opto-mechanics, where direct measurements of the quantisation are unavailable or technically difficult.

\begin{center}\begin{tabular*}{.5\textwidth}{c}\hline\end{tabular*}\end{center}

We thank E. Huntington for useful discussions.
This research was conducted by the {\it Australian Research Council Centre of Excellence for Quantum Computation and Communication Technology} (project number CE110001029). B.H. appreciates the support by the Alexander von Humboldt-Foundation.

\end{document}